\documentclass[12pt]{iopart}

\usepackage{iopams}  
\usepackage{graphicx}

\begin{document}
\title{When a negative weak value $-1$ plays the counterpart of a probability $1$}

\author{Kazuhiro Yokota  and Nobuyuki Imoto}
\address{Department of Materials Engineering Science,
Graduate School of Engineering Science, Osaka University,
Toyonaka, Osaka 560-8531, Japan}
\ead{yokota@qi.mp.es.osaka-u.ac.jp}

\date{\today}

\begin{abstract}
When the weak value of a projector is $1$, a quantum system behaves as in that eigenstate with probability $1$.
By definition, however, the weak value may take an anomalous value lying outside the range of probability like $-1$.
From the viewpoint of a physical effect, we show that such a negative weak value of $-1$ can be regarded as the counterpart of the ordinary value of $1$.
Using photons, we experimentally verify it as the symmetrical shift in polarization depending on the weak value given by pre-postselection of the path state.
Unlike observation of a weak value as an ensemble average via weak measurements, the effect of a weak value is definitely confirmed in Hong-Ou-Mandel effect: the symmetrical shift corresponding to the weak value can be directly observed as the rotation angle of a half wave plate.
\end{abstract}

\pacs{03.65.Ta, 42.50.Xa}

\maketitle
\section{Introduction}
\begin{figure}[b]
 \begin{center}
	\includegraphics[scale=1.0]{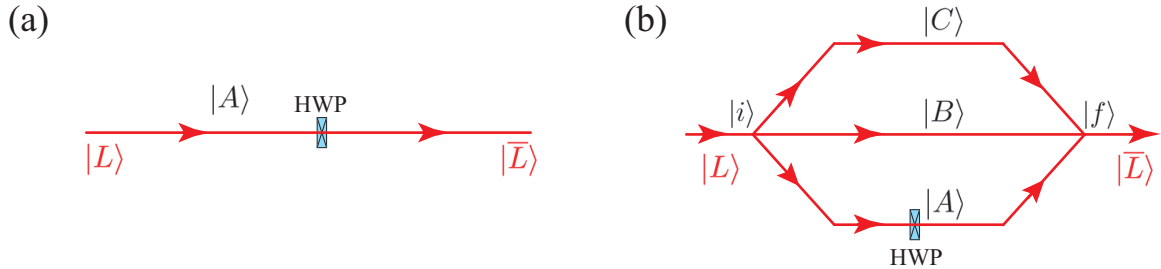}
 \end{center}
	\caption{(a)A photon certainly passes $|A\rangle$ and the polarization flips due to the HWP. (b)A photon is in superposition of the paths $|A\rangle$, $|B\rangle$, and $|C\rangle$.
When the path state of a photon is appropriately pre-postselected, the polarization flips as if it is operated by the HWP in $|A\rangle$.
}
\label{fig:3-box}
\end{figure}

In general it is difficult to determine the trajectory of a quantum particle, because a quantum particle can be in superposition of trajectories.
However, the particle in such a superposition can behave as if it takes a certain trajectory by choosing both initial and final states appropriately, namely, pre-postselection.
Fig.\ref{fig:3-box} (a) represents a case in which a photon takes only the path of $|A\rangle$ without superposition.
On the path, a half wave plate (HWP) is placed so that the polarization of a photon is flipped as $H\leftrightarrow V$, where $H(V)$ represents a horizontal (vertical) polarization (i.e. the angle of the HWP is $\pi/4$).
If an incident photon is, for simplicity, linearly polarized as $|L\rangle =\cos\theta|H\rangle +\sin\theta|V\rangle$, it is transformed into $|\overline{L}\rangle =\sin\theta|H\rangle +\cos\theta|V\rangle$.
Without loss of generality, we also assume $\sin\theta\le\cos\theta$ ($0\le\theta\le\pi /4$) hereafter.
On the other hand, in Fig.\ref{fig:3-box} (b) the path state of a photon is initially in superposition as $|i\rangle = (|A\rangle +|B\rangle +|C\rangle )/\sqrt{3}$;
with a finite probability, it is postselected in $|f\rangle = (|A\rangle +|B\rangle - |C\rangle )/\sqrt{3}$.
Suppose that the polarization is also initially in $|L\rangle$ and the HWP is set only on $|A\rangle$, too.
Then, after the postselection, the polarization of the photon also turns out to be $|\overline{L}\rangle$ as if it has passed $|A\rangle$ with certainty.
In fact such an application of pre-postselection was proposed by Y.Aharonov and L.Vaidman, who showed that one shutter can close $N$ slits with certainty \cite{PP1}.

In pre-postselection a time-symmetric formalism is often available.
According to the ABL formula \cite{TS1, TS2}, the probability to find a photon in $|A\rangle$ in the middle of the pre-postselection is given as follows,
\begin{eqnarray}
P(A) = \frac{|\langle f|A\rangle\langle A|i\rangle|^2}{|\langle f|A\rangle\langle A|i\rangle |^2+|\langle f|({\rm I}-|A\rangle\langle A|)|i\rangle |^2} = 1, \label{eq:ABL}
\end{eqnarray}
where ${\rm I}$ represents the identity operator.
The probability given by Eq.(\ref{eq:ABL}) means that a photon certainly passes $|A\rangle$.
In this case the polarization of a photon contains which-path information; observation of polarization corresponds to measuring whether a photon has passed $|A\rangle$ or not.
In fact $G\equiv\cos^2\theta-\sin^2\theta$, which satisfies $0\le G\le 1$ ($0\le\theta\le\pi /4$), can be regarded as the correlation (or measurement) strength between polarization and path.
When $G=1$ ($\theta =0$), the polarization has the complete which-path information:
if the polarization is found to be $|V\rangle$, we can confidently assert that the photon has passed $|A\rangle$.
On the other hand, we will conclude that the photon has not passed $|A\rangle$, if the polarization is $|H\rangle$.
The ABL formula is applied in such `strong' correlation;
the pre-postselected photon absolutely results in $|V\rangle$, agreeing with the probability of Eq.(\ref{eq:ABL}).

Meanwhile, when $G\sim 0$ ($\theta\sim\pi /4$), we cannot distinguish which path the photon has passed.
In this case of `weak ' correlation, a formalism other than the ABL formula should be applied, which has been known as weak value.
Weak value was introduced as a result of weak measurement without disturbance on a quantum state \cite{W1, W2}.
Given both initial and final states of $|i\rangle$ and $|f\rangle$, the weak value of an observable $\hat{O}$ is defined by $\langle\hat{O}\rangle_{\bf w} = \langle f|\hat{O}|i\rangle/\langle f|i\rangle$.
In this case the weak value of the projector $|A\rangle\langle A|$ is given by,
\begin{eqnarray}
p_A=\langle|A\rangle\langle A|\rangle_{\bf w} = \langle f|A\rangle\langle A|i\rangle /\langle f|i\rangle = 1.
\end{eqnarray}
It also seems to be reasonable that we interpret this value as probability as in Eq.(\ref{eq:ABL}): the pre-postselected photon certainly results in $|\overline{L}\rangle$ as passing through $|A\rangle$ with probability $1$.
Differently from the ABL formula, however, a weak value may lie outside the range of eigenvalue spectra.
In fact we can easily find $\langle |C\rangle\langle C|\rangle_{\bf w}=-1$, which cannot be regarded as a probability.

The example in Fig.\ref{fig:3-box} (b) has been known as the quantum box problem \cite{TS2, WP1};
such an anomalous weak value of $-1$ plays an important role in a quantum paradox \cite{WP1, WP2, WP3} (see also \cite{WP4, WP5}).
If we put the HWP on $|B\rangle$ instead of $|A\rangle$, the polarization results in $|\overline{L}\rangle$ too.
The story is the same as for $|A\rangle$:
as $\langle |B\rangle\langle B|\rangle_{\bf w}=1$, a photon behaves as if it has certainly passed $|B\rangle$.
However, it seems to be paradoxical that both $\langle |A\rangle \langle A|\rangle_{\bf w}=1$ and $\langle |B\rangle \langle B |\rangle_{\bf w}=1$, if we interpret them as probabilities.
Then the anomalous value of $\langle |C\rangle\langle C |\rangle_{\bf w}=-1$ is needed to hold the consistency as one: $\langle |A\rangle\langle A|\rangle_{\bf w}+\langle |B\rangle\langle B|\rangle_{\bf w}+\langle |C\rangle\langle C|\rangle_{\bf w}=1$ like conventional probabilities.

Does the anomalous value of $-1$ just arise to balance the numbers in a quantum paradox?
Does the value of $-1$ have any association with a physical operation, as the weak value of $1$ agrees with the probability $1$?
How is the polarization changed, when we put the HWP on $|C\rangle$ in Fig.\ref{fig:3-box} (b)?

In this paper, we discuss how the weak value of a projector can emerge in actuality and be associated with a physical operation.
As we have referred, we consider how the linear-polarization of a photon is changed by pre-postselection on the path state.
As a result, we show that the weak value $-1$ provides the shift of the angle of the polarization symmetrical to that one given by the weak value $1$.
We also demonstrated their symmetrical effects as actual angles of half wave plates directly by means of Hong-Ou-Mandel effect (two photon interference).

\begin{figure}
 \begin{center}
	\includegraphics[scale=1.0]{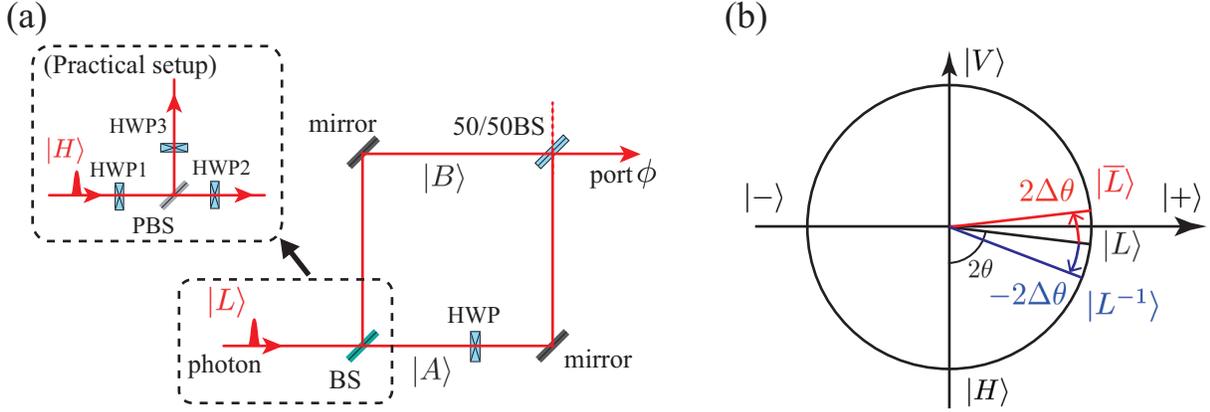}
 \end{center}
	\caption{(a)An interferometer to provide an arbitrary weak value, $\langle |A\rangle\langle A|\rangle_{\bf w}=p_A$.
In our experiment, PBS (polarized beam splitter) was substituted for BS as shown in the dashed box to prepare an expected initial state:
by adjusting HWP1, a preselection of the path state, $|\psi\rangle$, was achieved.
HWP2 and HWP3 were to make the polarization state of $|L\rangle$ in both $|A\rangle$ and $|B\rangle$.
Then the photon is in $|\psi\rangle|L\rangle$.
(b)The shift angles of polarization corresponding to the weak values of $p_A=1$ and $p_A=-1$ in Poincare sphere.
}
\label{fig:MZ}
\end{figure}

\section{Theory}
While we have given an example with three paths in Fig.\ref{fig:3-box} (b), the two paths as shown in Fig.\ref{fig:MZ} (a) are enough to prepare an arbitrary weak value, $\langle |A\rangle\langle A|\rangle _{\bf w}= p_A$.
For that purpose we choose the path states as follows:
After arriving at the beam splitter (BS) with an appropriate transmissivity/reflectivity, an incident photon evolves into superposition of paths: $|\psi\rangle =(p_A|A\rangle+(p_A-1)|B\rangle)/\sqrt{n}$ where $n$ is normalization.
Each path length is adjusted so that a photon is postselected in $|\phi\rangle=(|A\rangle-|B\rangle)/\sqrt{2}$, when a photon comes out from one of the port of 50/50BS (a beam splitter with reflectivity equal to transmissivity), $\phi$.
The weak value with this pre-postselection turns out to be $\langle |A\rangle\langle A|\rangle_{\bf w} = p_A$.
The HWP is attached only on $|A\rangle$ as in Fig.\ref{fig:3-box}.
Eventually the polarization of a photon at the port $\phi$ is given as follows:
$|\psi\rangle|L\rangle \ 
\rightarrow \ [p_A|A\rangle|\overline{L}\rangle+(p_A-1)|B\rangle|L\rangle]/\sqrt{n} \
\rightarrow \ [p_A|\overline{L}\rangle-(p_A-1)|L\rangle]/\sqrt{n'} \
= \ [(p_A\sin\theta -(p_A-1)\cos\theta)|H\rangle + (p_A\cos\theta -(p_A-1)\sin\theta)|V\rangle]/\sqrt{n'}$
with normalization $n'$.
As a result, the direction of the linear-polarization is changed depending on the weak value, $\langle|A\rangle\langle A|\rangle_{\bf w}=p_A$.
In fact, when $p_A=1$, the polarization results in  $|\overline{L}\rangle$ as in Fig.\ref{fig:3-box}.

To illustrate the shift of the direction, we show the $x-z$ plane of Poincare sphere in Fig.\ref{fig:MZ} (b).
The poles correspond to $|H\rangle$ and $|V\rangle$, and $|\pm\rangle=(|H\rangle \pm|V\rangle)/\sqrt{2}$ are in $x$-axis.
When $p_A=1$, as the polarizations of $|H\rangle$ and $|V\rangle$ are inverted each other, the polarization is given by reflection about $x$-axis to be $|\overline{L}\rangle$.
If we define the shift angle, $\Delta\theta$, as $|\overline{L}\rangle=\cos(\theta+\Delta\theta)|H\rangle +\sin(\theta+\Delta\theta)|V\rangle$ (i.e. $\Delta\theta = \pi /2-2\theta$), it is represented by $2\Delta\theta$ in Fig.\ref{fig:MZ} (b).
When $G=\cos^2\theta-\sin^2\theta\sim 0$ ($\theta\sim\pi /4$), we can easily approximate the shift angle as $\Delta\theta=G$.

On the other hand, when $p_A=-1$, the polarization results in $|L^{-1}\rangle =[(-\sin\theta+2\cos\theta)|H\rangle +(-\cos\theta+2\sin\theta)|V\rangle]/\sqrt{n'}$.
When $G\sim 0$, we can easily find that the shift angle is approximately given by $-\Delta\theta=-G$ (i.e. $-2\Delta\theta$ in  Fig.\ref{fig:MZ} (b)), which is symmetrical to that one by $p_A=1$.
Generally the shift angle is estimated as ${\rm Arctan}[(p_A\sqrt{1+G}-(p_A-1)\sqrt{1-G})/(p_A\sqrt{1-G}-(p_A-1)\sqrt{1+G})]-{\rm Arctan}\sqrt{(1-G)/(1+G)}\sim p_AG$ ($G\sim 0$).
In this sense positive and negative weak values affect the polarization symmetrically.

\begin{figure}
 \begin{center}
	\includegraphics[scale=1.0]{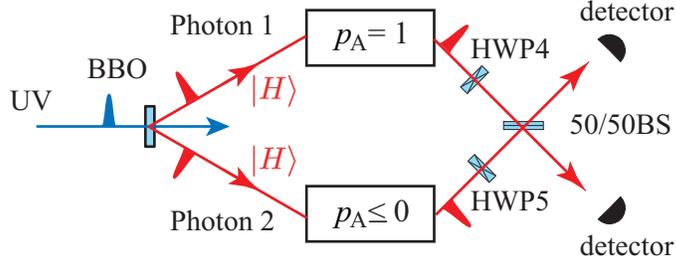}
 \end{center}
	\caption{Schematic of experimental setup to observe the symmetrical shifts in polarizations corresponding to the weak values of $p_A=1$ and $p_A=-1$.
Each box represents the interferometer in Fig.\ref{fig:MZ} (a) to prepare $p_A$.
While $|B\rangle$ can be omitted for $p_A=1$ (Photon 1), the interferometer for $p_A\le 0$ (Photon 2) achieved the visibility of $99.5\pm0.3\%$ with photons in $(|H\rangle +|V\rangle)/\sqrt{2}$.
Horizontally polarized photon pairs were generated via spontaneous parametric down-conversion from type I phase matched BBO crystal pumped by a UV pulse (a central wavelength of 395nm and an average power of 180mW).
The UV pulse is taken from the frequency-doubled Ti:sapphire laser (wavelength of 790nm, pulse width of 100fs, and repetition rate of 80MHz).
After passing the photon pair through HWP4 and HWP5, we observed the visibility of Hong-Ou-Mandel effect at 50/50BS.
}
\label{fig:exp}
\end{figure}

\section{Method}
To experimentally verify the symmetrical angles, we assembled the setup shown in Fig.\ref{fig:exp}.
We prepared a photon pair, one of which was pre-postselected for the weak values of $p_A=1$ (Photon 1), while the other one was for a negative weak value, $p_A\le 0$ (Photon 2).
As we have shown, the direction of linear-polarization of each photon was changed depending on each weak value.
We consider how these polarizations can be restored to the initial polarization, $|L\rangle$, by using half wave plates. 
The shift angle of polarization of Photon 1 is $\Delta\theta$ because of $p_A=1$, and the state results in $|\overline{L}\rangle$;
if the angle of HWP4 is $-\Delta\theta /2$ against the direction of $|\overline{L}\rangle$ (i.e. the angle is $\theta+\Delta\theta/2$ in real-space), the direction of polarization re-shifts by $-\Delta\theta$ and gets back into $|L\rangle$.
Correspondingly Photon 2 with $p_A\le 0$ can also return to $|L\rangle$ by HWP5.
In particular, when $p_A=-1$, the angle of HWP5 against $|L^{-1}\rangle$ should be $\Delta\theta/2$ (i.e. $\theta-\Delta\theta/2$), which is symmetrical to that one for $p_A=1$ against the initial direction of $|L\rangle$, namely, $\theta$.
To verify whether these photons are restored to the same state of $|L\rangle$, we observe the visibility of Hong-Ou-Mandel effect at 50/50BS:
we inquired about the negative weak value when the maximum visibility, which is ideally $1$, was obtained with the angle of HWP5 symmetrical to that one of HWP4.

\section{Result}
\begin{figure}
 \begin{center}
	\includegraphics[scale=0.7]{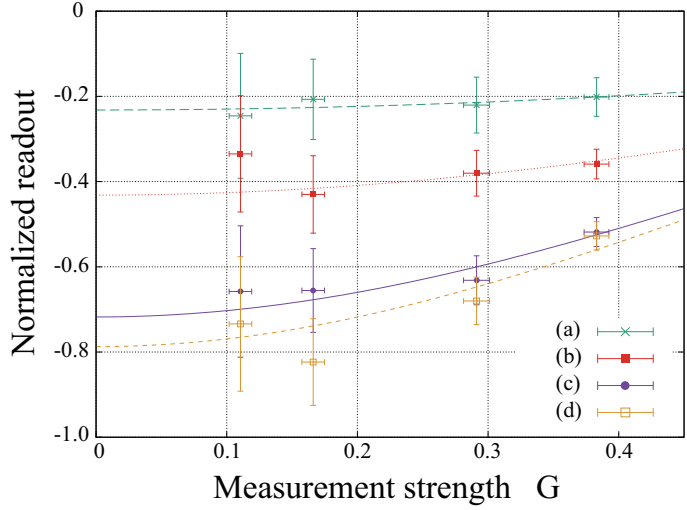}
 \end{center}
	\caption{The result of weak measurement on negative weak values.
Normalized readouts, $R(V|\phi)$, approach the weak values, as $G \rightarrow 0$.
$P(V|\phi)$ in Eq.(\ref{eq:read}) was derived from the counting ratio in $H$ and $V$.
To correct the unexpected polarization dependency, we normalized the counting ratio by that one in $G=0$: we calibrated $R(V|\phi)$ in $G=0$.
We also determined $G$ experimentally by means of that $R(V|\phi)=1$ $(0)$ when $p_A=1$ $(0)$.
By doing so, we also adjusted HWP2 and HWP3 in Fig.\ref{fig:MZ} (a) to provide the same $G$ in both $|A\rangle$ and $|B\rangle$.
The error bars of $R(V|\phi)$ contain the statistical errors and the errors of $G$.
The curves were derived by fitting the experimental data; the values when G=0, which correspond to weak values we prepared, were (a)$p_A=-0.23$, (b)$p_A=-0.43$, (c)$p_A=-0.72$, and (d)$p_A=-0.79$.
}
\label{fig:wv}
\end{figure}
Varying the angle of HWP1 in Fig.\ref{fig:MZ} (a), we produced the preselections of the path state (i.e. $|\psi\rangle$) to prepare various negative weak values, $p_A\le 0$, for Photon 2.
We performed four cases of the negative weak values to cover the area which includes the symmetrical angle of HWP5 to observe the maximum visibility as seen later: these negative weak values were experimentally determined by weak measurement as shown in Fig.\ref{fig:wv} (a)-(d).
As was referred to earlier, the polarization of a photon contains which-path information of $|A\rangle\langle A|$.
Suppose a photon passes $|A\rangle$ with certainty in Fig.\ref{fig:MZ} (a).
Then the photon results in $|\overline{L}\rangle =\sin\theta|H\rangle+\cos\theta|V\rangle$;
the probability of detecting the photon in $|V\rangle$ is given by $P(V|\phi)=\cos^2\theta$, which is larger than $P(H|\phi)=\sin^2\theta$ for detection of $|H\rangle$.
The contrast of $G=\cos^2\theta-\sin\theta^2$ can be regarded as the measurement strength on $|A\rangle\langle A|$:
although the polarization contains the information about which path an output photon has passed, such discrimination is lost more as $G\rightarrow 0$.
Then the path and the polarization have eventually no correlation: observation of polarization never brings about disturbance on the path state, which achieves weak measurement.
If we define a normalized readout by
\begin{eqnarray}
	R(V|\phi)=[P(V|\phi)-\sin^2\theta]/(\cos^2\theta-\sin^2\theta),   \label{eq:read}
\end{eqnarray}
the readout shows $R(V|\phi)=1$ in any $G$, which represents that the photon certainly in $|A\rangle$ well.
In an arbitrary pre-postselection, the normalized readout shows the weak value as $R(V|\phi)\rightarrow{\rm Re}\langle|A\rangle\langle A|\rangle_{\bf w}$, when $G\rightarrow 0$ \cite{WP3, WM1, WM2}.
Note that, when $G=1$ (strong measurement), it agrees with the ABL formula in Eq.(\ref{eq:ABL}), that is, $R(V|\phi)=P(A)$.
Fig.\ref{fig:wv} shows our experimental result of the weak measurement on negative weak values we prepared.

\begin{figure}
 \begin{center}
	\includegraphics[scale=0.5]{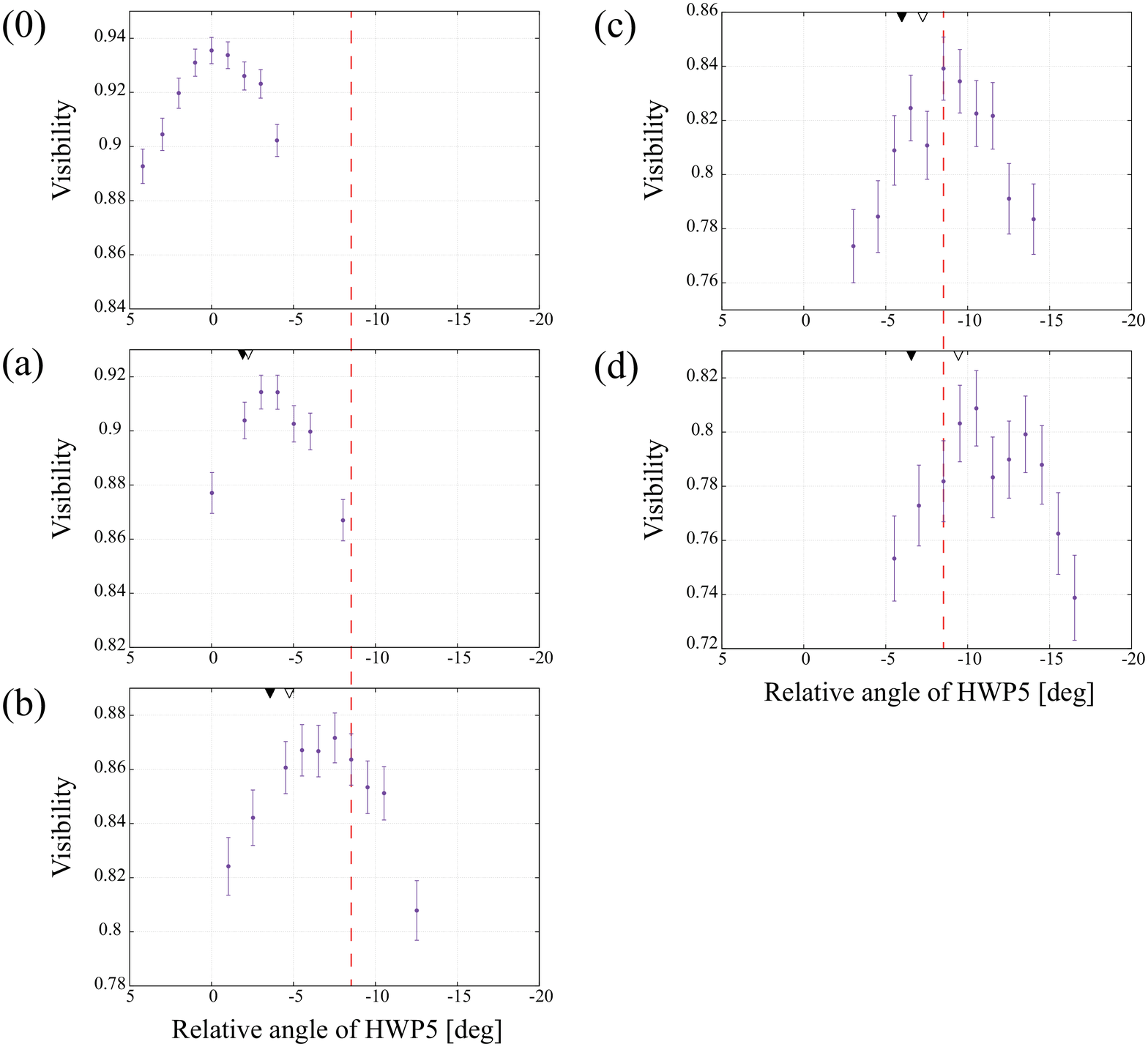}
 \end{center}
	\caption{Visibilities of Hong-Ou-Mandel effect between Photon 1 and Photon 2 when $G=0.29$.
(0) shows the result when the polarization of Photon 2 remained in $|L\rangle$ just before arriving at HWP5, namely, $p_A=0$.
(a)-(d) correspond to the negative weak values in Fig.\ref{fig:wv}.
The red dashed line indicates when the angle of HWP5 is set as $\theta-\Delta\theta/2$ (i.e. the relative angle is $-\Delta\theta/2$) which is symmetrical to the angle of HWP4 given by $\theta+\Delta\theta/2$.
According to the estimated weak value in Fig.\ref{fig:wv}, we also mark the theoretical angle, $p_A G/2$, for Photon 2 to be restored to $|L\rangle$ by using $\blacktriangledown$, while $\triangledown$ corresponds to the theoretical angle given by the expected weak value from the angle of HWP1 (see text).
}
\label{fig:vis}
\end{figure}

Using these weak values, we demonstrated the symmetrical shifts given by positive and negative weak values in $G=0.29\pm 0.01$.
The angle of HWP4 was, of course, $\pi/4$ so that Photon 1 with $p_A=1$ was restored to $|L\rangle$, because the polarization should be flipped as $H\leftrightarrow V$ again.
In fact, when $G=0.29$, we can easily calculate $\theta=36.57^\circ$ and $\Delta\theta=\pi/2-2\theta=16.86^\circ$, from which the angle of HWP4 is derived as $\theta+\Delta\theta /2=\pi/4$.
Varying the angle of HWP5, we observed the visibilities of Hong-Ou-Mandel effect between Photon 1 and Photon 2 as shown in Fig.\ref{fig:vis}:
we expect that the maximum visibility is observed with the angle of HWP5 of $\theta-\Delta\theta/2$, when the negative weak value is $p_A=-1$.
The high maximum visibility in Fig.\ref{fig:vis} $(0)$ shows that Photon 1 was certainly restored to $|L\rangle$, since Photon 2 stayed in $|L\rangle$.
Then the angle of HWP5 to achieve the maximum visibility corresponded to the direction of $|L\rangle$, which we define as $0$ degree in Fig.\ref{fig:vis}.
As the weak value became larger in negative (i.e. (a)$\rightarrow$(d)), the angle to achieve the maximum visibility shifted larger in response to $p_AG$ as mentioned previously.
The case of (c), in which the estimated weak value was about $p_A=-0.72$ in Fig.\ref{fig:wv}, gave almost the symmetrical angle to that one for $p_A=1$.
The slight disagreement with $p_A=-1$ mostly came from the noise counts.
As a weak value was larger in negative, the successful probability of postselection became smaller, by which the noise counts seemed to be larger relatively.
Actually the maximum visibility we achieved gradually decreased as (a)$\rightarrow$(d) in Fig.\ref{fig:vis}, although it must be ideally $1$.
As a result, a larger negative weak values were also estimated smaller than the values expected from the angle of HWP1 to prepare the preselection of $|\psi\rangle$, which were (a)$p_A=-0.27$, (b)$p_A=-0.57$, (c)$p_A=-0.87$, and (d)$p_A=-1.14$.
However, we were able to directly observe that the shift angle became larger, as the negative weak value was larger in negative;
the case of (c) gave the angle which was almost symmetrical to that one for Photon 1 (i.e. $p_A=1$), in which the negative weak value was nearly $-1$ (the estimated value of $-0.72$, the expected value of $-0.87$).

\section{Conclusion}
We have experimentally shown an actual effect given rise to by a weak value, which has rather been considered to be the statistical average of a huge number of  weak measurement results.
The polarization of a photon shifts depending on the weak value by pre-postselection on the path state.
In particular the shift angles corresponding to the weak values of $1$ and $-1$ are symmetrical against the initial direction of polarization.
We directly observed the symmetrical angles as the rotation angles of half wave plates with the aid of Hong-Ou-Mandel effect.

Weak value has been observed as an ensemble average via weak measurement so far, while weak value can also appear as a physical value in some cases \cite{W_ph1, W_ph2, step, gra_wv, W_ph3}.
In fact we have experimentally estimated weak values by the statistical procedure.
In our main experiment, however, we verified the effect of weak values as the actual angles of the half wave plates to restore polarizations.
Although we needed an ensemble to confirm the maximum visibility as is shown in Fig.\ref{fig:vis}, which looks like the shift of a pointer in weak measurement \cite{W1, WP1},
we can stay in the peek, once the angles of the half wave plates are set appropriately to given weak values:
in ideal case, a coincidence count never takes place due to Hong-Ou-Mandel effect, which is absolutely assured (not statistically but) with each photon pair.
In other words, we observed Hong-Ou-Mandel effect for each photon pair by rotating the angle of a half wave plate to the direction indicated by the weak value; in this sense, our approach will make weak value to be more real object.
That is why we observed Hong-Ou-Mandel effect to verify whether the polarizations were restored to the initial ones.

The weak value of $-1$ has been known as an important piece in a quantum paradox.
Although the value gives us a manner to treat the paradox consistently, it is hard to accept such a strange value as a conventional probability.
From a different perspective, we had the weak value be associated with an actual phenomenon.
We believe that our demonstration will be helpful to clarify the role of weak value in both foundation and application of quantum physics.

\section*{Acknowledgment}
On performing our experiment we thank Rikizo Ikuta, Toshiki Kobayashi and Takashi Yamamoto for their cooperations.
This work was supported by JSPS Grant-in-Aid for Scientific Research(A) JP16H02214.

\end{document}